\documentclass[12pt]{amsart}

\usepackage{amssymb}

\usepackage{euscript}
\usepackage{multicol}
\usepackage{amsmath}
\usepackage{times}

\newcommand{\tr}{\mbox{\upshape tr\ }}
\newtheorem{theorem}{Theorem}[section]

\numberwithin{equation}{section}

\title[On a sum rule for  Schr\"odinger operators]{ On a sum  rule for  Schr\"odinger operators with complex potentials}
\author{Oleg Safronov }

\email{osafrono@uncc.edu }
\thanks{The author  would like to thank B. Vainberg, S. Molchanov, A. Gordon, A. Laptev, R. Frank and P. Grigoriev for inspiring  and motivating discussions}


\begin{document}

\maketitle

\begin{abstract}
We study the distribution of  eigenvalues of the one-dimensional Schr\"odinger operator with a complex valued potential $V$. We prove that if $|V|$ decays faster than the Coulomb potential, then the series  of imaginary parts of square roots of  eigenvalues is convergent.
\end{abstract}

\section{Introduction}

 Let $V:[0,\infty)\mapsto {\Bbb  C}$   be a complex  valued  potential.
 The object of  our  investigation is  the one-dimensional Schr\"odinger operator 
$$
H=-\frac{d^2}{dx^2}+V(x)
$$
on the half-line with the Dirichlet  boundary condition at zero. Denote
by $\lambda_j$ the eigenvalues of   the operator $H$ lying outside of   the interval ${\Bbb R}_+=[0,\infty)$. Note, that the multiplicity of each eigenvalue  equals 1.

We  shall consider only potentials from the space  $L^1({\Bbb R}_+)$.
It is interesting, that in this  case, all non-real eigenvalues $\lambda$ of $H$ satisfy the estimate 
$$
|\lambda|\leq\Bigl(\int_0^\infty|V|dx\Bigr)^2.
$$
The proof of this result can be found in \cite{D1} (see  also \cite{AN}).
Recently, this result was (partially) generalized  to the multi-dimensional  case.
It was proven in \cite{S}, that  the condition  $|V|\leq C(1+|x|)^{-q}$  with $q>1$ implies that all non-real eigenvalues of $-\Delta+V$ are situated in a disk of a finite radius. However, the  estimate
$$
|\lambda|\leq C\Bigl(\int_{{\Bbb R}^d}(1+|x|)^{1-d}|V|dx\Bigr)^2
$$
has not been proven.

The paper \cite{FLLS} treats   the multi-dimensional case. ( Everywhere below, $\Re z$ and $\Im z$ denote the real and the imaginary parts of $z$.)
The one-dimensional version  of the  main result of \cite{FLLS} tells us,  that for any $t>0$, 
the eigenvalues $\lambda_j$ of $H$
lying outside the   sector $\{\lambda:\ \ |\Im \lambda|<t\ \Re \lambda\}$ satisfy the  estimate 
\begin{equation}\label{LT}
\sum |\lambda_j|^\gamma\leq C\int |V(x)|^{\gamma+1/2}dx,\qquad \gamma\geq 1,
\end{equation}
where  the constant  $C$  depends on $t$ and $\gamma$ (see also \cite{LT} for the case  when $V$ is real). 

Finally, we would like to mention the paper \cite{LS}.  It  deals with the natural question that arises in relation to the main  result of \cite{FLLS}: what estimates are valid for the eigenvalues  situated inside  the conical sector  $\{\lambda:\ |\Im \lambda|<t \Re \lambda\}$, where  the eigenvalues might be close to the positive half-line?
Theorems of the article \cite{LS} provide some  information about the
 rate of accumulation of  eigenvalues to the set ${\Bbb R}_+=[0,\infty)$. 
Namely, \cite{LS} gives  sufficient conditions on 
$V$ that  guarantee convergence of the sum
\begin{equation}\label{LapSa}
\sum_{a<\Re \lambda_j<b}|\Im \lambda_j|^\gamma<\infty
\end{equation}
for  $0\leq a<b<\infty$.

\bigskip

Both exponents $\gamma$ in \eqref{LT} and in \eqref{LapSa} are  not  less  than 1.
We suggest a method  that allows one to study  the case $\gamma=1/2$.

\begin{theorem}\label{main} Let
$V:{\Bbb R}_+\mapsto {\Bbb  C}$ satisfy the condition 
$$
\int_0^\infty
(1+|x|^p)|V(x)|dx<\infty,$$ for some $p\in(0,1)$.
Then
$$
\sum_j |\Im \sqrt{\lambda_j}|\leq C\Bigl(\int_0^\infty  |x|^p|V(x)|\,dx\bigl( \int_0^\infty  |V(x)|\,dx\bigr)^p+\int_0^\infty  |V(x)|\,dx\Bigr),
$$
where the positive constant $C$ depends on $p$, but  is independent of $V$.
\end{theorem}
The proof  is based on   the   so called trace formulas approach, which was also used in several papers \cite{DHK}, \cite{GK} and \cite{HK} to estimate the eigenvalue sums. We  would like to remark that in order to prove  Theorem~\ref{main}
one  not only needs to modify  the existing trace formula, but one  also needs  to estimate s-numbers of the compact operator $\sqrt{|V|}(-d^2/dx^2-\lambda\mp i0)^{-1}\sqrt{|V|}$ for real $\lambda$.

\section{ Proof of Theorem~\ref{main}}  

{\it 1.} Before  proving the theorem we  will acquaint  the reader with our notations.
As  it was already mentioned $\Re z$ and $\Im z$ denote the real and the imaginary parts of $z$. The class of compact operators $T$ having the property
$$
||T||_{{\frak S}_q}^q:=\tr (T^*T)^{q/2}<\infty,\qquad q\geq1,
$$
is called the Neumann-Schatten class ${\frak S}_q$. The functional 
$||T||_{{\frak S}_q}$ is a norm on ${\frak S}_q$.
For $T\in {\frak S}_1$ one can introduce
$
\det(I+T)
$ as the product of  eigenvalues of $I+T$. Note  that
$$
|\det(I+T)|\leq \exp(||T||_{{\frak S}_1}).
$$
Besides $\det(I+T)$, one can introduce the second  determinant by setting
$$
\det\, _2(I+T)=\det(I+T)e^{-\tr T}.
$$
The advantage of this definition is illustrated by the estimate
$$
|\det\, _2(I+T)|\leq \exp(C||T||^2_{{\frak S}_2}).
$$

{\it 2.}
The   basic   tool  of the proof   is  the trace  formula involving   the eigenvalues $\lambda_j$ and  the  perturbation determinant $\det (I+VR(z))$ where $R(z)=(-d^2/dx^2-z)^{-1}$.
  It  is  known that  the  eigenvalues   of the operator $H$ are  zeros of the  function $d(z)=\det (I+VR(z))$. Traditionally, one writes $z$  in the form $z=k^2$
and one considers the function $a(k)=d(k^2)$ with $k\in {\Bbb C}_+$ instead of $d(z)$.

Denote  by $k_j$ the zeros of the function $a(k)$ lying in the upper half-plane 
${\Bbb C}_+$.
We construct the Blaschke product $B(k)$ having the same   zeros as $a(k)$
$$
B(k)=\prod_j
\frac{k-k_j}{k-\overline{k_j}}\frac{k_j}{|k_j|}.
$$ 
It is  pretty  obvious  that  the ratio $a(k)/B(k)$ does not have zeros and therefore the function $\log(a(k)/B(k))$ is  well defined in the upper half-plane. Moreover, the ratio $a(k)/B(k)$ has the nice property  that 
$$
\Bigl|\frac{a(k)}{B(k)}\Bigr|=|a(k)|\qquad {\rm if }\,\,  k\in {\Bbb R}.
$$
The trace formula is a relation   that involves  an integral of the function $\log|a(k)|$ and the zeros $k_j$. The Blaschke product allows one to  separate   the contribution of  zeros into  the trace formula from other contributions. Indeed, since
$$
\log B(k)=\log(\prod_j \frac{k_j}{|k_j|})-2i\sum_j \frac{\Im k_j}{k}-i\sum_j \frac{\Im k_j^2}{k^2}-2i\sum_j\frac{\Im k_j^3}{3k^3}+O(k^{-4})
$$
as $k\to\infty,$ we obtain that the real part of the integral
$$
\int_{C_R}\log(B(k))\rho(k)dk,\qquad \rho(k)=(R^2-k^2),
$$
  over the contour, consisting of  the interval $[-R,R]$ and the half-circle of radius $R$, equals
  $$
  2\pi R^2\sum_j \Im k_j-\frac{2\pi}3\sum_j \Im k^3_j.
  $$
  for a sufficiently large $R>0.$
  It is also   clear  that
  $$
  \int_{C_R}\log\Bigl(\frac{a(k)}{B(k)}\Bigr)\rho(k)dk=0,
  $$
  since  the function $\log\Bigl(\frac{a(k)}{B(k)}\Bigr)$ is analytic in the upper half-plane. Thus, we obtain that
  $$
  \int_{C_R}\log(B(k))\rho(k)dk=
\int_{C_R}\log(a(k))\rho(k)dk,
  $$
which implies the  equality
$$
2\pi R^2\sum_j \Im k_j-\frac{2\pi}3\sum_j \Im k^3_j=\Re
\int_{C_R}\log(a(k))\rho(k)dk.
$$
Choose now $R=2\int|V|dx.$  We will shortly  see  how convenient  this choice is,
and  now we  will obtain an estimate of the quantity $\log(a(k))$.

 We  have to estimate  this quantity twice: first time, we have to estimate the absolute value $|\log(a(k))|$ under the condition that $|k|=R$; second time, we
 will establish  an upper estimate of  $\log|a(k)|$  on the interval $[-R,R]$.
 
 Let us carry out the computations  for $|k|= R$. 
The  arguments are borrowed  from  \cite{Kdiss}.
 Let us estimate  the derivative of the function $\psi(z)=\log a(k)$, $\, z=k^2$. We have 
 $$
 \psi'(z)=\tr (H-z)^{-1}V(-d^2/dx^2-z)^{-1}=$$
$$\sum_{j=0}^\infty(-1)^j
\tr\Bigl[(-d^2/dx^2-z)^{-1}WU(W(-d^2/dx^2-z)^{-1}WU)^jW(-d^2/dx^2-z)^{-1}\Bigr]
 $$
 where $U=V/|V|$ and $W=\sqrt{ |V|}$. Since, for $|k|\geq R$,
 $$|| W(-d^2/dx^2-z)^{-1}W ||\leq \frac{\int|V|dx}{|k|}\leq\frac12,$$
 we obtain that
 $$
 \Bigl| \psi'(z)\Bigr|\leq C\int|V|dx \int_{-\infty}^{\infty}\frac{d\xi}{|\xi^2-z|^2}\leq\frac{C_1\int|V|dx}{|\Im k|^{3}},\qquad k^2=z.
 $$
 Integrating   along the vertical line we will obtain that
 $$
 |\psi(z)|\leq \frac{C_0\int|V|dx}{|\Im k|}.
 $$
Consequently, 
for $\phi=Arg( k)$,
$$
|\psi(z)||\rho(k)|\leq \frac{C_0\int|V|dx}{|R\sin(\phi)|}|R^2(1-e^{i2\phi})|
\leq CR\int|V|dx
$$
on the circle $\{k:\,\, |k|=R,\, \Im k>0\}$. It implies the following  estimate for   the integral
$$
\Bigl|\int_{|k|=R,\\ \Im k>0}\log(a(k))\rho(k)dk\,\, \Bigr|\leq C\pi R^2 \int |V|dx.
$$

Assume  now that $k=\bar k$.  Let us estimate  the quantity 
$\log |a(k)|= \log |\det(I+VR(z))|$ from above. Due to the relation $\log |a(k)|= -\Re(\int V dx/ 2ik)+\log |\det_2(I+VR(z))|$, we  conclude  that
\begin{equation}
\label{WRW}
||WR(z)W||_{{\frak S}_2}\leq \frac{\int |V|dx }{|k|}\implies \log |a(k)|\leq C\Bigl(\frac{\int |V|dx }{|k|}+\bigl(\frac{\int |V|dx }{|k|}\bigr)^2\Bigr)
\end{equation} 
however this  estimate is not suitable for $k\to0.$ 
Therefore we  have to conduct our  reasoning  in  a more delicate way. 
Consider the  integral kernel of the operator $X=WR(z)W$.  It is a  function of the form
$$
cW(x)\int_{-\infty}^\infty
\frac{\sin(\xi x)\sin(\xi y)}{\xi^2-z}W(y)d\xi.$$
It follows clearly from this formula  that $X$ is representable as the integral
$$
X=c\int_{-\infty}^\infty
\frac{l^*_\xi l_\xi }{\xi^2-z}d\xi,
$$
where the linear  functional $l_\xi$ is defined by the relation
$$
l_\xi(u)=\int_{0}^\infty\sin(\xi y)W(y)u(y)dy
$$
and acts  from $L^2({\Bbb R}_+)$ to ${\Bbb C}$.

It is obvious that
$$
||l_\xi||^2\leq |\xi|^{p}\int |x|^{p} |V|\, dx,\quad 0<p<1.
$$
Moreover $||l_\xi-l_\eta||$ can be estimated in the following  way. Since
$$
|\sin(\xi y)-\sin(\eta y)|\leq 2|\sin\Bigl((\frac{\xi-\eta}2)y\Bigr)|\leq
 C|\xi-\eta|^{p/2}|y|^{p/2},
$$
we obtain that 
$$
||l_\xi-l_\eta||\leq C |\xi-\eta|^{p/2}\Bigl(\int  |x|^p|V(x)|\,dx\Bigr)^{1/2}.
$$
Consider now the operator $G_\xi=l^*_\xi l_\xi $. It is   clear  that
$$
||G_\xi||_{{\frak S}_1}\leq |\xi|^{p}\int |x|^{p} |V|\, dx,\quad 0<p<1. 
$$
Moreover, 
$$
||G_\xi-G_\eta||_{{\frak S}_1}\leq ||l_\xi-l_\eta||(||l_\xi||+||l_\eta||)\leq
$$
$$\leq C |\xi-\eta|^{p/2}(|\xi|^{p/2}+|\eta|^{p/2})\Bigl(\int  |x|^p|V(x)|\,dx\Bigr)
$$
Therefore  the following  representation  of the operator $X$ 
$$
X=c\Bigl(\int_{-\infty}^\infty
\frac{G_\xi -G_\eta }{\xi^2-z}d\xi +\frac {\pi i G_\eta} k\Bigr),\qquad  
\eta=|\Re z|^{1/2}
$$
implies that
$$
||X||_{{\frak S}_1}\leq C\Bigl(\int_{-\infty}^\infty \frac{|\xi-\eta|^{p/2}(|\xi|^{p/2}+|\eta|^{p/2})} {|\xi^2-\eta^2|}d\xi+ 
\frac {\eta^{p}}{|k|}\Bigr)\int_0^\infty  |x|^p|V(x)|\,dx.
$$
If $k\in {\Bbb R}$ is real , then we obtain that
\begin{equation}\label{X}
||X||_{{\frak S}_1}\leq \frac C{|k|^{1-p}}\int_0^\infty  |x|^p|V(x)|\,dx.
\end{equation}
Therefore,
$$
\int_{-R}^R\log |a(k)| \rho(k) dk\leq R^2\int_{-R}^R\log |a(k)|  dk\leq
$$
$$
R^2C\Bigl(\int_0^\infty  |x|^p|V(x)|\,dx\Bigr)\Bigl(\int_0^\infty  |V(x)|\,dx\Bigr)^p.
$$

Let us summarize  the results: we proved that
$$
 \sum_j \Im k_j-\frac{1}{3R^2}\sum_j \Im k^3_j\leq C\Bigl(\int_0^\infty  |x|^p|V(x)|\,dx\Bigl( \int_0^\infty  |V(x)|\,dx\Bigr)^p+\int_0^\infty  |V(x)|\,dx\Bigr).
$$
It  remains to notice that
$|k_j|\leq \int|V|dx=R/2$ implies 
$$
\frac{1}{3R^2}\Im k^3_j\leq \frac14 \Im k_j.
$$

\bigskip

\noindent
The proof is completed. $\,\,\,\,\,\,\, \Box $

\end{document}